\documentclass[seceqn]{elsart}
\usepackage{amssymb}
\usepackage{graphicx}

\journal{J. Math. Analysis and Appl.}

\begin{document}


\def\eqref#1{(\ref{#1})}
\def\eqrefs#1#2{(\ref{#1}) and~(\ref{#2})}
\def\eqsref#1#2{(\ref{#1}) to~(\ref{#2})}
\def\sysref#1#2{(\ref{#1})--(\ref{#2})}

\def\Eqref#1{Eq.~(\ref{#1})}
\def\Eqrefs#1#2{Eqs.~(\ref{#1}) and~(\ref{#2})}
\def\Eqsref#1#2{Eqs.~(\ref{#1}) to~(\ref{#2})}
\def\Sysref#1#2{Eqs. (\ref{#1}),~(\ref{#2})}

\def\secref#1{Sec.~\ref{#1}}
\def\secrefs#1#2{Sec.~\ref{#1} and~\ref{#2}}

\def\appref#1{Appendix~\ref{#1}}

\def\Ref#1{Ref.~\cite{#1}}
\def\Refs#1{Refs.~\cite{#1}}

\def\Cite#1{${\mathstrut}^{\cite{#1}}$}

\def\EQ{\begin{equation}}
\def\EQs{\begin{eqnarray}}
\def\endEQ{\end{equation}}
\def\endEQs{\end{eqnarray}}


\def\fewquad{\qquad\qquad}
\def\severalquad{\qquad\fewquad}
\def\manyquad{\qquad\severalquad}
\def\manymanyquad{\manyquad\manyquad}

\def\downupindices#1#2{{}^{}_{#1}{}_{}^{#2}}
\def\updownindices#1#2{{}_{}^{#1}{}^{}_{#2}}
\def\mixedindices#1#2{{\mathstrut}^{#1}_{#2}}
\def\downindex#1{{}_{#1}}
\def\upindex#1{{}^{#1}}

\def\eqtext#1{\hbox{\rm{#1}}}

\def\hp#1{\hphantom{#1}}

\def\Parder#1#2{
\mathchoice{\partial{#1} \over\partial{#2}}{\partial{#1}/\partial{#2}}{}{} }
\def\nParder#1#2#3{
\mathchoice{\partial^{#1}{#2} \over\partial{#3}^{#1}}{\partial^{#1}{#2}/\partial{#3}^{#1}}{}{} }
\def\mixedParder#1#2#3#4{
\mathchoice{\partial^{#1}{#2} \over\partial{#3}\partial{#4}}{\partial^{#1}{#2}/\partial{#3}\partial{#4}}{}{} }
\def\parder#1#2{\partial{#1}/\partial{#2}}
\def\nparder#1#2{\partial^{#1}/(\partial {#2})^{#1}}
\def\mixedparder#1#2#3#4{\partial^{#1}{#2}/\partial{#3}\partial{#4}}
\def\parderop#1{\partial/\partial{#1}}
\def\mixedparderop#1#2#3{\partial^{#1}/\partial{#2}\partial{#3}}

\def\u#1{u\upindex{#1}}
\def\deru#1{u\downindex{#1}}

\def\v#1{v\upindex{#1}}
\def\derv#1{v\downindex{#1}}
\def\xder#1{{#1}\downindex{x}}
\def\vder#1{{#1}\downindex{v}}
\def\derx#1{x\downindex{#1}}

\def\der#1{\partial_{#1}}
\def\nder#1#2{\partial^{#2}_{#1}}
\def\D#1{D\downindex{#1}}
\def\nD#1#2{D\mixedindices{#2}{#1}}
\def\Dt#1{D\mixedindices{#1}{t}}
\def\Dx#1{D\mixedindices{#1}{x}}
\def\coder#1{\partial^{#1}}
\def\coD#1{D\upindex{#1}}

\def\X#1{{\bf X}_{\rm #1.}}
\def\prX#1{{\bf X}_{\rm #1}^{(1)}}
\def\S#1{S_{\rm #1}}

\def\sgn{{\rm sgn}}

\def\G#1{{\mathcal G}_{#1}}

\def\ie/{i.e.}
\def\eg/{e.g.}
\def\etc/{etc.}
\def\const{{\rm const}}

\def\dash{------}

\begin{frontmatter}

\title{Exact solutions of semilinear radial wave equations in $n$ dimensions}

\author{Stephen C. Anco\corauthref{cor}}
\corauth[cor]{Corresponding author.}
\ead{sanco@brocku.ca}

\address{
Department of Mathematics,
Brock University, 
St. Catharines, ON L2S 3A1 Canada }

\author{Sheng Liu}

\address{
Department of Mathematics and Physics,
Petroleum University,
Beijing 102249, P.R. China}

\begin{abstract}
Exact solutions are derived for an $n$-dimensional radial wave equation
with a general power nonlinearity. 
The method, which is applicable more generally to other nonlinear PDEs,
involves an ansatz technique to solve 
a first-order PDE system of group-invariant variables
given by group foliations of the wave equation,
using the one-dimensional admitted point symmetry groups.
(These groups comprise scalings and time translations,
admitted for any nonlinearity power,
in addition to space-time inversions 
admitted for a particular conformal nonlinearity power). 
This is shown to yield not only group-invariant solutions
as derived by standard symmetry reduction, 
but also other exact solutions of a more general form.
In particular, 
solutions with interesting analytical behavior connected with blow ups 
as well as static monopoles 
are obtained. 
\end{abstract}


\end{frontmatter}

\section{ Introduction }

Exact solutions are of considerable interest in the analysis of 
nonlinear partial differential equations (PDEs) ---
in particular, similarity solutions and more general group-invariant solutions
typically provide insight into asymptotic behavior, 
critical or blow-up behavior,
and also give a means by which to test numerical solution methods. 
For a nonlinear PDE in two independent variables, 
invariance under a one-parameter point symmetry group leads to 
a reduction of order to a nonlinear ordinary differential equation (ODE)
for the corresponding group-invariant solutions 
\cite{Olver-book,BlumanAnco-book}.
The strong utility of this method lies in the fact that 
the point symmetries admitted by a given nonlinear PDE can be found,
often in explicit form, by an algorithmic procedure
(and nowadays there are many symbolic software programs available for
computation of point symmetries). 
As a consequence, all point-symmetry reductions to nonlinear ODEs
are readily derived for a given nonlinear PDE. 
However, it remains necessary to solve these ODEs to obtain 
the group-invariant solutions explicitly \cite{integrable}
and in most cases this is still very difficult.
Indeed, in general, it may be necessary to resort to special ansatzes
or ad hoc solution techniques in order to find explicit solutions. 

In this paper we explore an alternative method that is 
founded on group-invariant foliation equations \cite{Ovsiannikov}
to find exact solutions of nonlinear PDEs.
To set up the method, for any admitted point symmetry, 
the first step involves converting the given nonlinear PDE 
into an equivalent first-order PDE system 
whose independent and dependent variables are given by, respectively, 
the classical invariants and differential invariants of 
the point symmetry transformation. 
This yields what is known as a {\it group-resolving system}, 
since formally it amounts to constructing a group foliation 
of the nonlinear PDE by employing jet space coordinates 
adapted to the given point symmetry group. 
Viewed geometrically, 
this system quotients out the action of the given point symmetry group,
so the points in its solution jet space 
correspond (generally in a one-to-one fashion) 
to the orbits of the symmetry group 
in the solution jet space of the given nonlinear PDE \cite{orbits}. 
Each orbit is determined by integration of the system of defining equations
for the differential invariants of this symmetry group. 
These equations have a well known {\it automorphic property} 
\cite{SheftelWinternitz}
that all of their solutions are related through a transitive group action 
by symmetry transformations,
allowing the equations to be explicitly integrated 
once any particular solution is known. 
Hence the problem of finding exact solutions of the original nonlinear PDE 
is reduced just to seeking particular explicit solutions of 
the group-resolving system. 
As the next step, 
the key idea of our method is to solve the group-resolving system 
by a separation of variables ansatz technique 
that is tailored to the form of nonlinearities in the PDEs of this system, 
expressed in terms of the group-invariant variables.  
In practice, with this technique, many explicit solutions are easily found,
whose form would not be readily obvious just by trying simple ansatzes 
using the original independent and dependent variables
in the nonlinear PDE,
or by simply writing down the form for classical group-invariant solutions. 
Most important, 
because all solutions of the nonlinear PDE are contained in 
each of its group-resolving systems, 
the method is not limited to obtaining exact solutions 
that just arise from classical point-symmetry reduction. 

We will apply this systematic method to 
a physically and analytically interesting 
$n$-dimensional radial wave equation of a semilinear form 
with a general power nonlinearity. 
In \secref{pointsymm}, 
the point symmetries and corresponding group-invariant reductions 
for this wave equation are considered. 
We find that the nonlinear ODEs given by the reductions 
admit extra point symmetries in some cases,
leading to reduction to quadrature for certain group-invariant solutions
of the wave equation. 
These solutions in particular include 
finite energy monopoles and static ``solitons'', 
which are of analytical interest. 
In \secref{method},
the method using group-resolving systems associated with
the admitted point symmetries of the wave equation is introduced,
employing our ansatz technique. 
Exact solutions obtained from this method are summarized in \secref{results},
along with a discussion of their analytical features of interest. 
Our results give several new families of 
time-dependent solutions for $n>1$,
including some with analytical properties that pertain to 
blow-up phenomena in $n\geq 2$ dimensions. 
(Additional results for the case of $n=1$ dimensions 
will be reported elsewhere.)
Finally, some concluding remarks are made in \secref{remarks}.

\section{ Point-symmetry reductions }
\label{pointsymm}

We consider the semilinear radial wave equation
$-\deru{tt}+\deru{rr} +\frac{n-1}{r}\deru{r} = \pm u |u|^{q-1}$
for $u(t,r)$, with a nonlinear interaction described by a power $q\neq 1$,
where $r$ denotes the radial coordinate in $n>1$ dimensions. 
Global (long-time) behavior of solutions to the Cauchy problem 
for this wave equation has attracted much analytical interest
over the past few decades (see \Refs{Strauss,Shatah} for a summary).
Note that stability of solutions 
relies on the conserved energy 
$\mathcal{E} =\int_0^\infty( 
\frac{1}{2}\deru{t}{}^2 +\frac{1}{2}\deru{r}{}^2 
\pm \frac{1}{q+1}|u|^{q+1} ) r^{n-1} \d r$
to be positive definite, 
which requires the sign of the nonlinear interaction term 
in the wave equation to be positive. 
In this case, 
it is known that any smooth initial data $u(0,r),\deru{t}(0,r)$
evolves to a smooth global solution $u(t,r)$ for all $t>0$
provided the nonlinearity power $q$ is less than a critical value 
\EQ
q_*=(n+2)/(n-2)=1+4/(n-2)
\endEQ
(in particular this includes all powers $q<\infty$ when $n=2$). 
The same result has been proved to hold in the critical case 
$q_*=5$ when $n=3$. 
For supercritical powers $q>q_*$ when $n\ge 3$, 
it is widely expected based on numerical studies that 
some smooth initial data will evolve to a solution $u(t,r)$ that blows up
at a finite time $t=T<\infty$. 
In light of these results, 
exact solutions of this wave equation may be of definite 
analytical and numerical interest. 

For our purpose it will be useful to work with 
a slightly modified nonlinear interaction term, 
\EQ\label{waveeq}
\deru{tt} -\deru{rr} -(n-1)r^{-1}\deru{r} = k \u{q} ,\quad
k=\pm 1 ,
\endEQ
which is easier to handle for finding exact solutions. 
Note, here, we will regard $u$ to be a real-valued function,
as is compatible with the case of all integer values $q$;
in the case of non-integer values $q$, we will either 
restrict the domain of $u(t,r)$ so that $u>0$, 
or allow a formal complexification of $u(t,r)$.
In any case, 
it will be straightforward to patch together such solutions, if necessary, 
to satisfy the more analytically well-behaved wave equation 
$-\deru{tt}+\deru{rr} +\frac{n-1}{r}\deru{r} = \pm u |u|^{q-1}$.

The geometrical point symmetries admitted by the wave equation \eqref{waveeq}
for $n>1$ are well known \cite{Strauss}:
\EQs
&&\eqtext{ time translation } \quad
\X{trans} =\parderop{t} \quad\eqtext{ for all $q$},
\label{transsymm}\\
&&\eqtext{ scaling } \quad
\X{scal} =t\parderop{t} + r\parderop{r} +\frac{2}{1-q} u\parderop{u} 
\quad\eqtext{ for all $q\neq 1$},
\label{scalsymm}\\
&&\eqtext{ space-time inversion } \quad
\X{inver} =(t^2+r^2)\parderop{t} + 2rt\parderop{r} +(1-n)t u\parderop{u} 
\nonumber\\ &&\manyquad\severalquad
\eqtext{ only for $q=(n+3)/(n-1)$},
\label{inversymm}
\endEQs
where $\X{}$ is the infinitesimal generator of a one-parameter group of
point transformations acting on $(t,r,u)$. 
The special power for which the inversion exists is commonly called 
the conformal power
\EQ
q_c=(n+3)/(n-1) = 1+4/(n-1) .
\endEQ


To begin, we discuss the group-invariant solutions arising
from reduction of the wave equation \eqref{waveeq} 
to a nonlinear ODE under each of these point symmetries.
It will be worth pointing out that these ODEs each inherit 
a variational structure related to the natural action principle 
\EQ\label{action}
\S{}[u] =\int_{-\infty}^{+\infty} \int_0^\infty (
-\deru{t}{}^2 +\deru{r}{}^2 - \frac{2k}{q+1}\u{q+1} ) r^{n-1} \d r\d t 
\endEQ
for this wave equation. 
Note that, as we easily see, 
this action principle is formally invariant under both 
the time translation and inversion symmetries, for any dimension $n$,
but not under the scaling symmetry 
unless the dimension is $n=1+4/(q-1)$ 
corresponding to the conformal power $q=q_c$.

\subsection{ Time translation invariance }

The form for group-invariant solutions is obviously $u=U(r)$,
satisfying the ODE
\EQ\label{transode}
U''+(n-1) r^{-1} U' +k U^q =0 . 
\endEQ
Despite its simplicity, 
this nonlinear second-order ODE is intractable to solve explicitly in general.
However, for special values of the parameters $n$ and $q$, 
it will be possible to integrate \eqref{transode} completely to quadrature
if it admits a variational point symmetry (see \Ref{Olver-book} for details),
\ie/ a symmetry of ODE \eqref{transode} that leaves invariant 
the reduced action principle 
\EQ
\S{trans.}[U] = \int_0^\infty ( U'{}^2 -\frac{2k}{q+1} U^{q+1} ) r^{n-1} \d r
\endEQ
for this ODE.
The computation of point symmetries is straightforward 
and leads to the following result.

Proposition~1:
For translation-invariant solutions of wave equation \eqref{waveeq},
the point symmetries admitted by ODE \eqref{transode} for $n>1$
are comprised by
\EQ\label{transXscal}
\eqtext{ scaling } \quad
\X{scal} =r\parderop{r} +\frac{2}{1-q} U\parderop{U} 
\quad\eqtext{ for all $q\neq 1$},
\endEQ
and 
\EQs
&&\eqtext{ non-rigid dilation } \quad
\X{dil} =r^{2-n}( r\parderop{r} +(n-2) U\parderop{U} )
\nonumber\\ &&\severalquad\fewquad
\eqtext{ only for $q=(4-n)/(n-2)$, $n\neq 2$}. 
\label{transXdil}
\endEQs
The dilation is a variational symmetry for all $n\neq 2$
(\ie/ such that $q\neq\infty$), 
while the scaling is only a variational symmetry 
when $n= 2(q+1)/(q-1) \neq 2$
corresponding to the critical power $q=1+4/(n-2)=q_*$. 

The appearance of special powers, especially the critical power,  
is interesting because these powers are not distinguished by
the geometrical point symmetries of the wave equation. 
Note that the non-rigid dilation generalizes a uniform dilation group, 
$r \rightarrow \lambda r$, $u\rightarrow \Omega u$, 
such that $\lambda,\Omega$ are no longer constants 
but depend on a power of $r$. 

To carry out the reduction to quadrature implied by Proposition~1, 
we change variables to canonical coordinates 
derived from the coefficients of the infinitesimal transformations $\X{}$
(see \Ref{BlumanAnco-book} for details)
in the variational case. 
For the scaling symmetry \eqref{transXscal},
these coordinates are given by 
$x=\ln r$, $v=r^{-1+n/2} U$, 
yielding the following ODE for $v(x)$
\EQ\label{transvscaleq}
v'' -p^2 v +k v^q =0 ,\quad
p=1-n/2,\ q=1+4/(n-2)
\endEQ
(\ie/ a nonlinear oscillator equation)
which is easily integrated to quadrature. 
Thus, we have the implicit general solution $(n\neq 2)$
\EQ\label{transvscalsol}
x+\tilde c = 
\pm \int 1/\sqrt{ 2c + (1-n/2)^2 v^2 -k(1-2/n) v^{2n/(n-2)} } dv 
\endEQ
with arbitrary constants $c,\tilde c$. 
This includes as a particular case 
the invariant solution $v=\const$ of reduced ODE \eqref{transvscaleq}
under the scaling symmetry \eqref{transXscal},
given explicitly by 
\EQ\label{transvinvscalsol}
v= ( (1-n/2)/\sqrt{k} )^{n/2 -1} . 
\endEQ

Canonical coordinates for the dilation symmetry \eqref{transXdil}
are given by $x=\frac{1}{n-2} r^{n-2}$, $v=r^{n-2} U$. 
This yields a similar integrable ODE for $v(x)$
\EQ\label{transvdileq}
v'' +k v^q =0 ,\quad
q=-1+2/(n-2)
\endEQ
which has the implicit general solution $(n\neq 2)$
\EQ\label{transvdilsol}
x+\tilde c = \pm \int 1/\sqrt{ 2c -k(n-2) v^{2/(n-2)} } dv 
\endEQ
with arbitrary constants $c,\tilde c$. 
We note, here, 
the invariant solution of reduced ODE \eqref{transvdileq}
under the dilation symmetry \eqref{transXdil} is trivial, 
$v=0$. 

Thus, symmetry reduction yields two translation-invariant solutions,
in implicit general form \eqrefs{transvscalsol}{transvdilsol},
for special powers $q=1+4/(n-2) =q_*$ and $q=-1+2/(n-2)$, respectively.
In the case of other powers, 
although the ODE \eqref{transode} 
admits the scaling symmetry \eqref{transXscal} for all $q\neq 1$, 
and therefore its order can be reduced through a symmetry reduction procedure
\cite{Olver-book,BlumanAnco-book}, 
the reduced first-order ODE is not tractable to solve 
by any standard integrating factor methods \cite{BlumanAnco-book},
apart from the invariant case analogous to 
the solution \eqref{transvinvscalsol}. 
Consequently, it would appear to be necessary to try special ansatzes
or ad hoc techniques to solve the second-order ODE \eqref{transode} 
to obtain additional translation-invariant solutions. 

A similar situation occurs for the other group-invariant solutions,
as we will now see. 

\subsection{ Scaling invariance }

The form for group-invariant solutions is easily seen to be $u=t^p U(\xi)$,
where $p=2/(1-q)$ and $\xi=r/t$,
with $U(\xi)$ satisfying the  nonlinear second-order ODE
\EQ\label{scalode}
(1-\xi^2)U''+((n-1)\xi^{-1}+2(p-1)\xi)U' +p(1-p)U +k U^q =0 . 
\endEQ
This ODE is considerably more complicated than 
the one for translation-invariant solutions. 
Nevertheless, 
it will still be possible to reduce \eqref{scalode} to quadrature
for special values of the parameters $n$ and $q$ 
if it admits a variational point symmetry. 
An added complication here is that the action principle \eqref{action}
does not have a well-defined scaling reduction
(due to its lack of invariance under the scaling symmetry),
but this difficulty can be by-passed if we consider multipliers $h(\xi)$
for the ODE \eqref{scalode} (see \Ref{BlumanAnco-book}). 
We find that $h(\xi)= (1-\xi^2)^{-p+(1-n)/2}$ leads to the existence of
an action principle 
\EQ
\S{scal.}[U] = \int_{-\infty}^{+\infty} ( 
p(1-p) U^2-(1-\xi^2) U'{}^2+\frac{2k}{q+1} U^{q+1} ) 
h(\xi) \xi^{n-1} \d\xi . 
\endEQ
Now, a straightforward computation of point symmetries 
gives the following result.

Proposition~2:
For scaling-invariant solutions of wave equation \eqref{waveeq},
the point symmetries admitted by ODE \eqref{scalode} consist of only
\EQs
&&\eqtext{ non-rigid dilation } \quad
\X{dil} =\sqrt{|1-\xi^2|} ( \xi\parderop{\xi} +(1-n/2) U\parderop{U} ) 
\nonumber\\ &&\manyquad\fewquad
\eqtext{ for $q=(n+2)/(n-2)$, $n\neq 2$}. 
\label{scalXdil}
\endEQs
This symmetry is variational for $n= 2(q+1)/(q-1)\neq 2$
corresponding to the critical power $q=1+4/(n-2)=q_*$ for all $n\neq 2$.

Thus, ODE \eqref{scalode} has a reduction to quadrature
precisely in the case of the critical power. 
Changing variables to canonical coordinates, given by 
$v=\xi^{-1+n/2} U$, 
$x=\frac{1}{2}\ln\frac{1-\sqrt{1-\xi^2}}{1+\sqrt{1-\xi^2}}$ if $|\xi|\leq 1$
and $x=\frac{\pi}{2}-\arctan\frac{1}{\sqrt{\xi^2-1}}$ if $|\xi|\geq 1$
(with $x$ smooth at $\xi=1$), 
we obtain the following ODE for $v(x)$
\EQ\label{scalvdileq}
s v'' -p^2 v +k v^q =0 ,\quad
p=1-n/2, q=1+4/(n-2)
\endEQ
where $s=\sgn(1-\xi^2)$. 
This nonlinear ODE has the same form 
as ODE \eqref{transvscaleq}, 
with the implicit general solution $(n\neq 2)$
\EQ\label{scalvdilsol}
x+\tilde c = 
\pm \int 1/\sqrt{ 2c +s (1-n/2)^2 v^2 -s k(1-2/n) v^{2n/(n-2)} } dv 
\endEQ
with arbitrary constants $c,\tilde c$. 
As a special case, 
the invariant solution $v=\const$ of reduced ODE \eqref{scalvdileq}
under the dilation symmetry \eqref{scalXdil} 
is given by the same expression \eqref{transvinvscalsol} 
as for ODE \eqref{transvscaleq}. 

Hence, other than the special case \eqref{transvinvscalsol} 
and its obvious generalization to non-critical powers, 
symmetry reduction yields just the scaling-invariant solution
in implicit general form \eqref{scalvdilsol},
for the critical power $q=1+4/(n-2) =q_*$. 

\subsection{ Conformal invariance }

The form for group-invariant solutions here is given by $u=r^p U(\xi)$
with $p=2/(1-q) =(1-n)/2$ and $q=q_c=1+4/(n-1)$
where $\xi=(t^2-r^2)/r$ is the invariant associated with inversions
(radial conformal transformations) on the space and time variables. 
The ODE satisfied by $U(\xi)$ is readily found to be 
\EQ\label{inverode}
\xi^2 U''+ 2\xi U' -p(p+1)U +k U^{q} =0 . 
\endEQ
In contrast with the scaling case, 
it is straightforward to derive an action principle 
\EQ
\S{inver.}[U] = \int_{-\infty}^{+\infty} ( 
p(p+1) U^2 + \xi^2 U'{}^2 -\frac{2k}{q+1} U^{q+1} ) \d\xi
\endEQ
for the nonlinear second-order ODE \eqref{inverode},
without need for a multiplier,
due to the dilation invariance of the action principle \eqref{action}. 
Proceeding as before, we now determine the variational point symmetries
admitted by this ODE. 

Proposition~3:
For conformally-invariant solutions of wave equation \eqref{waveeq},
the only point symmetry admitted by ODE \eqref{inverode} for $n>1$ is
a non-variational scaling of its independent variable
\EQ\label{dilXscal}
\eqtext{ scaling } \quad
\X{scal} = \xi\parderop{\xi} . 
\endEQ

Because this scaling is not a variational symmetry,
we do not obtain a complete reduction of the ODE \eqref{inverode}
to quadrature. 
In particular, with a change of variables to canonical coordinates 
$x=\ln\xi$, $v=U$, 
the ODE becomes
\EQ\label{dilvscaleq}
v'' +v'-p(p+1) v +k v^{q} =0 ,\quad
p=(1-n)/2, q=1+4/(n-1)
\endEQ
(\ie/ a damped nonlinear oscillator equation).
This ODE is intractable to solve explicitly in general, 
except in the case of the invariant solution $v=\const$ 
under the scaling symmetry \eqref{dilXscal}, 
which yields
\EQ\label{invervinvdilsol}
v= ( (n-1)(n-3)/(2\sqrt{k}) )^{(n-1)/4} . 
\endEQ
However, without resorting to special ansatzes or other techniques,
we do not obtain any other explicit conformally-invariant solutions
through symmetry reduction. 

\subsection{ Group action on solutions
and invariance under optimal subgroups }
\label{groupaction}

We now discuss some deeper group-theoretical aspects of 
our results on invariant solutions of wave equation \eqref{waveeq}.
First it is natural to look at how these solutions transform
under the full symmetry group, $\G{q}$, 
generated by the geometrical point symmetries 
\eqref{transsymm}, \eqref{scalsymm}, \eqref{inversymm}. 
Note the group structure depends on the nonlinearity power $q\neq 1$.
The one-dimensional point symmetry subgroups of 
translations, scalings, and inversions 
are shown in the Appendix.

For all $q\neq 1$ other than the conformal power $q_c$, 
the point symmetry algebra is given by the commutator of 
the translation and scaling generators
\EQ
[\X{trans},\X{scal}] = \X{trans} .
\label{transscalcomm}
\endEQ
This generates a solvable two-dimensional symmetry group
$\G{q\neq q_c} \simeq U(1) \rtimes U(1)$
consisting of a semidirect product of 
the scaling subgroup \eqref{scalgroup}
and the translation subgroup \eqref{transgroup}. 
The explicit families of 
translation-invariant solutions \eqrefs{transvscalsol}{transvdilsol}
found for special powers of $q$ 
each remain invariant under the action of this group,
subject to a shift 
$(\tilde c,c) \rightarrow (\tilde c +\ln\lambda, c)$
and a scaling 
$(\tilde c,c) \rightarrow (\tilde c \lambda^{2-n}, c \lambda^{(4-2n)/(n-3)})$
on the integration constants. 
In contrast, the explicit family of 
scaling-invariant solutions \eqref{scalvdilsol} is non-invariant,
as it gives rise to a one-dimensional orbit 
under the time translation subgroup,
with the integration constants remaining unchanged. 

For the case of the conformal power $q=q_c$,
the point symmetry algebra is enlarged by commutators 
involving the inversion generator, 
\EQ
[\X{trans},\X{inver}] = 2\X{scal} ,\quad
[\X{scal},\X{inver}] = \X{inver} .
\label{invercomm}
\endEQ
The commutators \eqrefs{transscalcomm}{invercomm} together 
generate a semisimple three-dimensional symmetry group
$\G{q=q_c} \simeq SL(2,\mathbb R)$
obtained by composition of the inversion subgroup \eqref{invergroup}
with the scaling and translation subgroups \eqrefs{scalgroup}{transgroup}. 
Under the action of the full group,
the single explicit solution \eqref{invervinvdilsol}
found for the conformal power $q_c$ is invariant. 

An interesting feature arises from the three-dimensional nature of
the symmetry group structure in the conformal power case. 
With respect to conjugation by the full group,
the translation and inversion subgroups fall into the same conjugacy class,
while the scaling subgroup comprises its own conjugacy class.
The origin of this structure is easily explained by means of 
the following special conformal point symmetry 
of wave equation \eqref{waveeq},
\EQ
u(t,r) \rightarrow (t^2-r^2)^{-p} u(-t/(t^2-r^2),r/(t^2-r^2)) ,\quad 
p=2/(1-q_c) 
\label{involution}
\endEQ
which corresponds to an inner automorphism of 
the symmetry group $SL(2,\mathbb R)$, 
as shown in the Appendix. 
This transformation acts as an involution 
$\X{trans} \leftrightarrow -\X{inver}$, 
$\X{scal} \leftrightarrow -\X{scal}$,
on the symmetry algebra. 
Thus it follows that a complete enumeration of conjugacy classes 
of the symmetry group 
is represented by the one-dimensional subalgebras
$\X{trans}$ or $\X{inver}$, $\X{scal}$, and $\X{trans}+\X{inver}$. 
These are referred to as the optimal subalgebras 
for group-invariance considerations \cite{Winternitz,Olver-book}. 

As a consequence, 
through the point symmetry transformation \eqref{involution},
ODE \eqref{transode} for translation-invariant solutions with $q=q_c$
is equivalent to ODE \eqref{inverode} for conformally-invariant solutions.
An essentially different ODE will arise for 
solutions invariant under the subgroup generated by 
\EQ
\X{trans} + \X{inver} 
=(1+ t^2+r^2)\parderop{t} + 2rt\parderop{r} +(1-n)t u\parderop{u} 
\label{transinversymm}
\endEQ
combining translations and inversions,
as we now investigate. 

The form for such group-invariant solutions is 
a generalization of conformally-invariant solutions, 
given by  
$u=r^p U(\xi)$ where, now, $\xi= (1+t^2-r^2)/r$,
with $p=2/(1-q)=1-n/2$ and $q=q_c=1+4/(n-1)$. 
The corresponding reduction of wave equation \eqref{waveeq} 
yields the ODE
\EQ\label{transinverode}
(\xi^2 +4)U''+ 2\xi U' -p(p+1)U +k U^{q} =0 .
\endEQ
Compared to the conformal case,
scaling invariance is obviously lost in this reduction. 
A computation of point symmetries for this ODE 
shows that the extra dilation invariance holding in the translation case
is also lost. 

Proposition~4:
ODE \eqref{transinverode} 
for group-invariant solutions of wave equation \eqref{waveeq} 
admits no point symmetries for $n>1$. 

This result is perhaps unsurprising 
because of the less geometric nature of
the symmetry group generator \eqref{transinversymm} here. 
The lack of invariance of ODE \eqref{transinverode} 
means it has no symmetry reductions. 
It is also intractable to solve by integrating factor methods 
\cite{BlumanAnco-book},
and hence we do not find any explicit solutions.

\section{ Group foliation method and ansatzes }
\label{method}

The basis of our method is the introduction of 
group foliation equations 
associated with the point symmetries admitted by 
the wave equation \eqref{waveeq}.
There is an algorithmic construction for group foliations \cite{Ovsiannikov}
in terms of group-invariant jet space coordinates 
given by the invariants and differential invariants of
an admitted one-dimensional point symmetry group. 
To explain the ideas, 
we first consider the scaling symmetry \eqref{scalsymm}. 

A convenient choice of invariants and differential invariants is given by 
\EQ
x=t/r ,\quad
v=r^{-p} u
\label{scalxv}
\endEQ
and 
\EQ
G=r^{1-p} \deru{t} ,\quad
H=r^{1-p} \deru{r} 
\label{scalGH}
\endEQ
where $p=2/(1-q)$. 
While $x$ and $v$ here are mutually independent, 
$G$ and $H$ are related by equality of mixed $r,t$ derivatives 
on $\deru{t}$ and $\deru{r}$, which gives
\EQ
\D{r}( r^{p-1} G ) = \D{t}( r^{p-1} H ) 
\label{GHmixedeq}
\endEQ
where $\D{r}, \D{t}$ denote total derivatives with respect to $r,t$. 
Furthermore, 
$v,G,H$ are necessarily related through the wave equation \eqref{waveeq} by
\EQ
\D{t}( r^{p-1} G ) -\D{r}( r^{p-1} H ) = r^{p-2}( (n-1)H + k\v{q} ) . 
\label{GHwaveeq}
\endEQ
Now we put $G=G(x,v)$, $H=H(x,v)$ into equations \eqrefs{GHmixedeq}{GHwaveeq}
to arrive at a first-order PDE system
\EQs
(p-1)G -x\xder{G} +(H-pv) \vder{G} -\xder{H} -G\vder{H} =0 , 
\label{scalGHeq1}\\
\xder{G} + G\vder{G} -(p+n-2)H+x\xder{H} -(H-pv)\vder{H} =k\v{q} , 
\label{scalGHeq2}
\endEQs
with independent variables $x,v$, and dependent variables $G,H$. 
These PDEs will be called the {\it scaling-group resolving system}
for the wave equation \eqref{waveeq}. 

We establish, first of all, that this system \sysref{scalGHeq1}{scalGHeq2}
contains all scaling-invariant solutions of the wave equation \eqref{waveeq}.
Consider any solution 
\EQ\label{scalu}
u(t,r)= r^p U(x) ,
\endEQ
with scaling-invariant form,
where
\EQ\label{scalUeq}
(1-x^2) U'' + (2p+n-3)x U' -p(p+n-2) U -k U^q =0
\endEQ
is the ODE given by reduction of PDE \eqref{waveeq}. 
From relation \eqref{scalGH} we have
\EQ
G=U' ,\quad
H=p U-x U' , 
\label{GHUrel}
\endEQ
which together yield 
\EQ
H=p v-x G . 
\label{scalinvGHsol}
\endEQ
This relation is easily verified to satisfy PDE \eqref{scalGHeq1}. 
In addition, PDE \eqref{scalGHeq2} simplifies to 
\EQ
(1-x^2)( \xder{G} + G\vder{G} ) + (2p+n-3)x G  = p(p+n-2) v +k \v{q} . 
\label{scalinvGHeq}
\endEQ
We then see that the characteristic ODEs for solving this first-order PDE
are precisely 
\EQs
&& dv/dx =G ,\quad
(1-x^2) dG/dx + (2p+n-3)x G =p(p+n-2) v +k v^q ,
\nonumber\\&&
\endEQs
which are satisfied due to equations \eqrefs{scalUeq}{GHUrel}.
Hence, we have proved the following result.

Lemma~1:
Scaling-invariant solutions \eqref{scalu} of wave equation \eqref{waveeq}
correspond to solutions of 
the first-order PDE system \sysref{scalGHeq1}{scalGHeq2}
with the specific form \eqref{scalinvGHsol}. 

We now demonstrate that, in fact, a larger correspondence holds.

Lemma~2:
There is a mapping \eqrefs{scalxv}{scalGH} between 
the solutions $u(t,r)$ of wave equation \eqref{waveeq}
and the solutions $G(x,v),H(x,v)$ of 
its scaling-group resolving system \sysref{scalGHeq1}{scalGHeq2}. 
The map $u \rightarrow (G,H)$ is many-to-one and onto;
its inverse $(G,H) \rightarrow u$ is one-to-many and onto. 

Proof: 
For any solution pair $G(x,v),H(x,v)$,
we note that the total derivative equation \eqref{GHmixedeq} holds
by PDE \eqref{scalGHeq1},
and hence $G$ and $H$ have the form \eqref{GHUrel} 
for some function $u(t,r)$ which, however, is non-unique. 
Through PDE \eqref{scalGHeq2}, 
we also have the total derivative equation \eqref{GHwaveeq}.
Substitution of expressions \eqref{GHUrel} into this equation 
then yields the wave equation \eqref{waveeq} on $u(t,r)$. 

To show the converse for any solution $u(t,r)$, 
we express $u$ as a function $v(t,x)=(x/t)^p u(t,t/x)$ 
through relation \eqref{scalxv}.
Without loss of generality, 
we will assume $\derv{t}$ is not identically zero,
as otherwise we have a group-invariant solution $v=v(x)$
for which Lemma~1 already establishes 
existence of a corresponding solution
$G(x,v),H(x,v)$ of the system \sysref{scalGHeq1}{scalGHeq2}.
By substitution of the resulting expressions \eqref{scalxv} 
into the relation \eqref{scalGH}, 
we obtain a pair of functions $G(t,x),H(t,x)$ that satisfy 
the total derivative equations \eqrefs{GHmixedeq}{GHwaveeq}.
Now we use the implicit function theorem to express $t=T(v,x)$
for some function $T$, 
which exists at least locally because, by assumption,
$\derv{t}\neq 0$ in some open domain in the $(t,x)$-plane. 
Elimination of $t$ in $G$ and $H$ then produces functions 
$G=g(x,v)=G(T(v,x),x)$, $H=h(x,v)=H(T(v,x),x)$. 
Since $G$ and $H$ satisfy equations \eqrefs{GHmixedeq}{GHwaveeq},
the pair $g(x,v),h(x,v)$ is a solution of 
system \sysref{scalGHeq1}{scalGHeq2}.
$\qed$

The many-to-one nature of this mapping arises when equation \eqref{scalGH}
is integrated to obtain $u(t,r)$ from $G(x,v),H(x,v)$. 
Through equation \eqref{scalxv}, 
the integration consists of solving a consistent pair of 
parametric first-order ODEs
\EQ
\deru{t} =r^{p-1} G(t/r,r^{-p} u) ,\quad
\deru{r} =r^{p-1} H(t/r,r^{-p} u) 
\label{u}
\endEQ
whose general solution will involve a single arbitrary constant. 

The PDE system \sysref{scalGHeq1}{scalGHeq2} for $G(x,v),H(x,v)$
is well-suited to solve by a separation of variables ansatz
if we exploit the feature that 
the nonlinearity in this system is only quadratic, 
while the interaction term 
given by a power nonlinearity $\u{q}$ in the wave equation 
now appears only as an inhomogeneous term $\v{q}$ in the system. 
Thus we will consider solutions of the form 
\EQ
G=g(x) \v{a} ,\quad
H=h(x) \v{b} 
\label{GHansatz}
\endEQ
with a power dependence on $v$. 
In outline the steps to find $a,b,g(x),h(x)$ proceed as follows. 
First the powers $a,b$ are determined by 
substituting expressions \eqref{GHansatz} 
into the PDEs \eqrefs{scalGHeq1}{scalGHeq2} and balancing powers of $v$, 
in particular, $\v{a},\v{b},\v{a+b-1}$ in PDE \eqref{scalGHeq1},
and $\v{a},\v{b},\v{2a-1},\v{2b-1},\v{q}$ in PDE \eqref{scalGHeq2}. 
This leads to a number of cases to consider for $a,b$, 
taking into account that the term $\v{q}$ must be balanced by 
at least one term among the other powers. 
Equating coefficients of like powers in each case 
then yields an overdetermined system of algebraic and differential equations
for $g(x),h(x)$. 
Their overdetermined nature allows all these systems to be readily solved. 
In those cases where nontrivial solutions are found, 
we are left with finding $u(t,r)$ by integration of 
a parametric ODE system \eqref{u},
which reduces to solving separable ODEs
\EQ
\u{-a} \deru{t} = r^{p(1-a)-1} g(t/r) ,\quad
\u{-b} \deru{r} = r^{p(1-b)-1} h(t/r) . 
\label{uodes}
\endEQ
Rather than consider all possible cases exhaustively, 
we will now go through in detail only those cases for which 
the system of equations obtained on $g(x),h(x)$ is 
found to be the least overdetermined,
giving the most promising chance for nontrivial solutions to exist. 
As suggested by compatibility of the pair of ODEs \eqref{uodes},
these cases are distinguished by the relation $a=b$. 

Case $a=b=(q+1)/2$: 
Here, PDE \eqref{scalGHeq1} involves just the power $\v{(q+1)/2}$,
whose coefficient then yields the ODE 
\EQ
xg'+h' =0 . 
\label{gode1}
\endEQ
PDE \eqref{scalGHeq2} contains two powers $\v{(q+1)/2},\v{q}$,
and their respective coefficients give 
the ODE
\EQ
g'+(1-n)h +x h' =0 , 
\label{hode1}
\endEQ
in addition to the algebraic equation
\EQ
g^2 -h^2 = 2k/(q+1) . 
\label{gheq1}
\endEQ
Differentiation of equation \eqref{gheq1} 
followed by substitution of equation \eqref{gode1} gives 
$(x h+g) g'=0$
and hence 
\EQ
g'=0
\quad \eqtext{ or }\quad 
xh+g=0 . 
\endEQ
For the latter possibility $g=-xh$, 
equation \eqref{hode1} reduces to $n h=0$,
which determines $h=0$. 
But this implies $g=-x h =0$, 
contradicting equation \eqref{gheq1}. 
Therefore we must have the other possibility, $g'=0$.
Equations \eqrefs{gode1}{hode1} together now yield $(n-1) h =0$
and hence, for $n>1$,
\EQ
h =0 . 
\label{hsol}
\endEQ
Then equation \eqref{gheq1} determines
\EQ
g=\pm \sqrt{ 2k/(q+1) } . 
\label{gsol}
\endEQ
Thus, we have a solution of PDE system \sysref{scalGHeq1}{scalGHeq2}. 
We remark that, in the less interesting case $n=1$,
the solution generalizes to 
$g=\pm \sqrt{ (c^2+2k)/(q+1) }$, 
$h=c/\sqrt{ (q+1) }$,
with an arbitrary constant $c$. 

Case $a=b=q$: 
PDE \eqref{scalGHeq1} again involves a single power, 
yielding the ODE 
\EQ
h' + xg' -g =0 . 
\label{gode2}
\endEQ
Likewise, PDE \eqref{scalGHeq2} again contains two powers, 
leading to 
the ODE
\EQ
xh' -n h +g' =k
\label{hode2}
\endEQ
and the algebraic equation
\EQ
g^2 =h^2 . 
\label{gheq2}
\endEQ
Clearly, from equation \eqref{gheq2} we have $g=\pm h$,
and then equations \eqrefs{gode1}{hode2}
reduce to an overdetermined pair of ODEs
\EQ
(x\pm 1) h' = k+n h =h . 
\endEQ
Elimination of $h'$ gives the algebraic equation $h=k/(1-n)$,
implying $h' =0$, which is inconsistent. 
Therefore, PDE system \sysref{scalGHeq1}{scalGHeq2} 
has {\it no} solution in the present case, for any $n\ge 1$. 

Unsurprisingly, 
all additional cases obtained from balancing other powers
can be shown to yield no solutions except if $n=1$. 

Because separation of variables \eqref{GHansatz} is successful 
in yielding solutions of PDE system \sysref{scalGHeq1}{scalGHeq2}, 
we will consider a more general two-term ansatz
\EQ
G=g_1(x) \v{a} + g_2(x) v ,\quad
H=h_1(x) \v{b} + h_2(x) v . 
\label{GHtwotermansatz}
\endEQ
The restriction to linear powers for the additional terms here
is motivated by the particular form of nonlinearities 
occurring in the PDEs. 
This ansatz \eqref{GHtwotermansatz} is found to yield three more solutions
for $n>1$. 

Proposition~5:
Through ansatzes \eqrefs{GHansatz}{GHtwotermansatz}, 
the scaling-group resolving system \sysref{scalGHeq1}{scalGHeq2}
for $n>1$ 
has the solutions: 
\EQs
G = \pm \sqrt{2k/(q+1)} \v{(q+1)/2} ,\quad
H = 0 , 
\label{scalingsol1}\\
G=0 ,\quad
H=(2-n) v \pm \sqrt{(2-n)k} \v{1/(n-2)} ,\quad
q=\frac{4-n}{n-2} , 
\label{scalingsol2}\\
G = \pm k (n-3)^{-1} \v{(n-1)/(n-2)} ,\quad
H = (2-n) v \pm G ,\quad
q=\frac{n-1}{n-2} , 
\label{scalingsol3}\\
G= \pm \sqrt{-k} \v{-1} + x^{-1} v ,\quad
H=0 ,\quad
q=-3 . 
\label{scalingsol4}
\endEQs

To continue, we next write down the group-resolving systems
associated with the translation and inversion point symmetries
for the wave equation \eqref{waveeq},
and present the solutions obtained via our ansatz technique. 

It should be noted that the inversion-group resolving system is equivalent to
the translation-group resolving system 
upon specialization to the case of the conformal power $q=q_c$, 
which follows from the involution property of 
the special conformal point symmetry transformation \eqref{involution}
admitted by wave equation \eqref{waveeq}.
However, the complexities of the ansatz technique 
differs for these two systems,
and actually as a result we will see different solutions emerge. 

Obvious invariants and differential invariants 
for the translation symmetry \eqref{transsymm} are simply 
\EQ
x=r ,\quad
v=u
\label{transxv}
\endEQ
and 
\EQ
G= \deru{t} ,\quad
H=\deru{r} . 
\label{transGH}
\endEQ
The resulting first-order PDE system is given by 
\EQs
\xder{G} +H\vder{G} -G\vder{H} =0 , 
\label{transGHeq1}\\
G\vder{G} -H\vder{H} -\xder{H} -(n-1) x^{-1} H =k\v{q} , 
\label{transGHeq2}
\endEQs
with independent variables $x,v$, and dependent variables $G,H$. 

Lemma~3:
Solutions of 
the translation-group resolving system \sysref{transGHeq1}{transGHeq2}
of the form
\EQ
G=0, \quad
H=U'(x)
\endEQ
correspond to translation-invariant solutions $u=U(x)$ 
of wave equation \eqref{waveeq}. 
More generally, 
the solutions $u(t,r)$ of the wave equation 
and the solutions $G(x,v),H(x,v)$ 
of its translation-group resolving system 
are related by a many-to-one, onto mapping $u \rightarrow (G,H)$,
and a one-to-many, onto inverse mapping $(G,H) \rightarrow u$,
through equations \eqrefs{transxv}{transGH}. 

Explicit solutions are readily found 
by the same separation of variables ansatzes introduced in the scaling case. 
However, up to a change of variable, 
these solutions are merely the same as 
the ones \eqref{scalingsol1}, \eqref{scalingsol2}, \eqref{scalingsol3} 
derived in the scaling case that 
have no dependence on the scaling invariant $x=t/r$. 
This reflects the fact that the PDE system \sysref{transGHeq1}{transGHeq2}
admits a scaling symmetry, 
leading to invariant solutions of the form 
\EQ
G(r,u) = r^{p-1} g(r^{-p} u) , \quad
H(r,u) = r^{p-1} h(r^{-p} u) ,\quad
p=2/(1-q)
\endEQ
where $g,h$ depend only on the scaling invariant $v=r^{-p} u$
and satisfy the scaling-group resolving system 
\sysref{scalGHeq1}{scalGHeq2}. 

For the inversion symmetry \eqref{inversymm},
invariants $x(t,r),v(u,t,r)$ are given through the equations 
\EQs
\X{inver} x = (t^2+r^2) \derx{t} +2rt \derx{r} =0 ,
\\
\X{inver} v = (1-n) tu \derv{u} + (t^2+r^2) \derv{t} +2rt \derv{r} =0 . 
\endEQs
Integration of the characteristic ODEs of these two equations yields 
\EQ
x=(t^2-r^2)/r ,\quad
v=r^{(n-1)/2} u . 
\label{inverxv}
\endEQ
Differential invariants 
$G(\deru{t},\deru{r},u,t,r),H(\deru{r},\deru{t},u,t,r)$,
with essential dependence on $\deru{t},\deru{r}$, respectively, 
are obtained in a similar way from solving the equations
\EQ
\prX{inver} G = 0 ,\quad
\prX{inver} H =0 , 
\endEQ
where 
\EQs
\prX{inver} =
\X{inver} 
+ ((1-n) u-(1+n) t\deru{t} -2r \deru{r} )\parderop{\deru{t}}
-( (1+n) t\deru{r} +2r \deru{t} )\parderop{\deru{r}} 
\nonumber\\ 
\endEQs
is the prolongation of the generator $\X{inver}$ from 
$(t,r,u)$ to $(t,r,u,\deru{t},\deru{r})$
(see \Refs{Olver-book,BlumanAnco-book}). 
This leads to the expressions
\EQs
G= r^{(n-1)/2} ( (t^2+r^2) \deru{t} +2rt \deru{r} +(n-1) tu ) , 
\label{inverG}\\
H= r^{(n-1)/2} ( (t^2+r^2) \deru{r} +2rt \deru{t} 
+(n-1)\frac{t^2+r^2}{2r} u ) .
\label{inverH}
\endEQs
The reciprocal relations
\EQs
\deru{t}= r^{(1-n)/2} (t^2 -r^2)^{-2}( (t^2+r^2) G -2rt H ) , 
\label{inverut}\\
\deru{r}= r^{(1-n)/2} (t^2 -r^2)^{-2}( (t^2+r^2) H -2rt G ) 
+\frac{1-n}{2r} u , 
\label{inverur}
\endEQs
give rise to the following first-order PDE system 
\EQs
x^2\xder{G} +G\vder{H} -H\vder{G} =0 , 
\label{inverGHeq1}\\
x^{-2} ( G\vder{G} -H\vder{H} ) +\xder{H} = -p(p+1) v + k\v{q} , 
\label{inverGHeq2}
\endEQs
with independent variables $x,v$, and dependent variables $G,H$,
where $q=(n+3)/(n-1)$ and $p=(1-n)/2=2/(1-q)$. 

Lemma~4:
Conformally-invariant solutions $u= r^{(n-1)/2} U(x)$ 
of wave equation \eqref{waveeq} 
correspond to 
solutions of conformal-group resolving system \sysref{inverGHeq1}{inverGHeq2}
of the form
\EQ
G=0, \quad
H=-x^2 U'(x) .
\endEQ
Through equations \eqref{inverxv}, \sysref{inverG}{inverur}, 
the solutions $u(t,r)$ of the wave equation 
and the solutions $G(x,v),H(x,v)$ 
of its conformal-group resolving system 
are related more generally by 
a many-to-one, onto mapping $u \rightarrow (G,H)$,
and a one-to-many, onto inverse mapping $(G,H) \rightarrow u$. 

The previous separation of variables ansatz 
again readily yields explicit solutions. 
In contrast to the translation case, 
the system \sysref{inverGHeq1}{inverGHeq2} 
here is able to yield new solutions. 

Proposition~6:
Through ansatzes \eqrefs{GHansatz}{GHtwotermansatz}, 
the conformal-group resolving system \sysref{inverGHeq1}{inverGHeq2}
for $n>1$ 
has the solution 
\EQ
G = \pm \sqrt{k (n-1)/(n+1)} x\v{(n+1)/(n-1)} ,\quad
H = \frac{1}{2} (n-1) x v . 
\label{inversol1}
\endEQ

\subsection{ Group-invariant potentials and ansatzes }

We introduce an extension of our ansatz technique by taking advantage of
conservation laws that arise for each of 
the group-resolving systems. 
To illustrate the ideas, we consider the translation case first. 

It is easily seen that each of the PDEs in the 
translation-group resolving system \sysref{transGHeq1}{transGHeq2}
admit a conservation law
\EQs
(H/G)_v = (-1/G)_x , 
\label{transconslaw1}\\
( x^{n-1}(G^2-H^2) )_v = ( 2 x^{n-1} H + \frac{2k}{n} x^{n} \v{q} )_x , 
\label{transconslaw2}
\endEQs
where, recall, $x=r$ and $v=u$ are the translation invariants. 
Associated with these respective conservation laws 
are potential variables (see \Ref{Bluman}) given by 
\EQs
\Phi_x = H/G ,\quad
\Phi_v = -1/G , 
\label{transpot1}\\
\Psi_x = x^{n-1}(G^2-H^2)/2 ,\quad
\Psi_v = x^{n-1} H + \frac{k}{n} x^{n} \v{q} , 
\label{transpot2}
\endEQs
with corresponding reciprocal relations
\EQ
G = -1/\Phi_v ,\quad 
H = -\Phi_x/\Phi_v , 
\endEQ
and 
\EQ
G = \pm x\sqrt{ 2x^{-1-n} \Psi_x +(x^{-n} \Psi_v  -\frac{k}{n} \v{q})^2 }
,\quad
H=  x^{1-n} \Psi_v  -\frac{k}{n} x \v{q} . 
\endEQ
The system of PDEs \sysref{transGHeq1}{transGHeq2}
can be rewritten in terms of each potential $\Phi,\Psi$:
\EQ
( (1-\Phi_x{}^2)/\Phi_v )_v = -2 x^{1-n} ( x^{n-1} \Phi_x/\Phi_v )_x +2k\v{q}
\label{transpot1eq}
\endEQ
and 
\EQs
\left( (x^{-n} \Psi_v  -\frac{k}{n} \v{q})/
\sqrt{ 2x^{-1-n} \Psi_x +(x^{-n} \Psi_v  -\frac{k}{n} \v{q})^2 } \right)_v 
\nonumber\\
= -\left( x^{-1}/ 
\sqrt{ 2x^{-1-n} \Psi_x +(x^{-n} \Psi_v  -\frac{k}{n} \v{q})^2 } \right)_x . 
\label{transpot2eq}
\endEQs
Note, equations \eqrefs{transpot1}{transpot2} each provide 
a one-to-many, onto embedding of solutions $G(x,v),H(x,v)$ of
the translation-group resolving system 
into solutions $\Phi(x,v),\Psi(x,v)$ of 
the respective associated potential systems. 

We remark that 
the inhomogeneous term in the conservation law \eqref{transconslaw2}
can be shifted between the two sides,
altering the potential $\Psi$ by a multiple of the corresponding term 
$\frac{2k}{n(q+1)} x^{n} \v{q+1}$. 
Throughout, we simply fix this freedom in the most convenient way
for solving the resulting potential systems. 

We will now employ a separation of variables ansatz 
\EQ
\Phi = \phi(x) v^a ,\quad
\Psi = \psi(x) v^b
\label{potansatz}
\endEQ
for finding explicit solutions of each PDE \eqrefs{transpot1eq}{transpot2eq}. 
As before, the steps consist of balancing powers of $v$ to determine $a,b$,
followed by solving an overdetermined system of 
algebraic and differential equations for $\phi(x)$ or $\psi(x)$. 
In particular, despite the complexity of their appearance, 
the form of these PDEs becomes quite manageable to handle 
after the substitution \eqref{potansatz} is carried out. 
This readily leads to the results
\EQs
\Phi = \pm \sqrt{ 2(q+1)/k } (q-1)^{-1} \v{(1-q)/2} , 
\\
G= \mp \sqrt{ 2k/(q+1) } \v{(q+1)/2} ,\quad
H =0 , 
\label{transpotoldsol}
\endEQs
and
\EQs
\Psi = \frac{1}{n( n(1-q)+1+q )} x^n \v{q+1} , 
\\
G= \pm \frac{1}{n(1-q)+1+q} \v{q} 
\sqrt{ x^2(q-1)^2 +2(n(1-q)+1+q) \v{1-q} } ,
\nonumber\\
H = \frac{q-1}{n(1-q)+1+q} x \v{q} . 
\label{transpotsol}
\endEQs
We note that the second solution \eqref{transpotsol} is new,
while the first solution \eqref{transpotoldsol} reproduces 
the solution \eqref{scalingsol1} already obtained in Proposition~5
directly from the translation-group resolving system. 
This duplication occurs because the form of $G,H$ for the $\Phi$ potential 
using the ansatz \eqref{potansatz} is precisely the same as
the original separation of variables ansatz \eqref{GHansatz}. 

The situation is similar in the scaling case as well as the conformal case.
For the conformal-group resolving system 
\sysref{inverGHeq1}{inverGHeq2}, 
there are conservation laws 
\EQs
( x^{-2} H/G )_v = ( 1/G )_x , 
\label{inverconslaw1}\\
( x^{-2} (G^2-H^2) )_v = -2( H +p(p+1) xv +k x \v{q} )_x , 
\label{inverconslaw2}
\endEQs
where  $x=(t^2-r^2)/r$ and $v=r^{(n-1)/2} u$ are the conformal invariants,
with $q=q_c=(n+3)/(n-1)$ and $p=(1-n)/2$. 
Using the potential 
\EQ
\Psi_x = (G^2-H^2)/x^2 ,\quad
\Psi_v = -2H-2x( p(p+1) v +k\v{q} )
\endEQ
associated with the conservation law \eqref{inverconslaw2}, 
we find the ansatz \eqref{potansatz} leads to no solutions;
however, via a two-term ansatz
\EQ
\Psi = \psi_1(x) v^b + \psi_2(x) v^2
\label{pottwotermansatz}
\endEQ
motivated by the form of the PDE for this potential,
we obtain 
\EQ
\Psi = kx \v{q+1} -p^2 x \v{2}
\label{inverpsi}
\endEQ
and
\EQ
\Psi = 2\frac{k}{q+1} x \v{q+1} -p^2 x \v{2} .
\label{inverpsi'}
\endEQ
The first result \eqref{inverpsi} 
gives a new solution
\EQ
G =\pm x\sqrt{ p^{-2} \v{2q} -k\v{q+1} } ,\quad
H = -p x v + k p^{-1} x\v{q} ,
q=\frac{n+3}{n-1} ,\quad p=\frac{1-n}{2} . 
\label{inverpotsol} 
\endEQ
The second result \eqref{inverpsi'} 
reproduces the solution already obtained in Proposition~6.
This same solution is the only one found through the other potential 
\EQ
\Phi_x = H/(x^2 G) ,\quad
\Phi_v = 1/G
\endEQ
with use of our ansatzes. 

The scaling-group resolving system \sysref{scalGHeq1}{scalGHeq2}
has conservation laws 
\EQs
( G/(H+ x G-p v) )_v = -( 1/(H+ x G-p v) )_x , 
\label{scalconslaw1}\\
( G+ x H )_x -( \frac{1}{2} (H^2-G^2) -p vH +\frac{k}{q+1} \v{q+1} )_v
= (2p+n-1) H , 
\label{scalconslaw2}
\endEQs
with $x=t/r$ and $v=r^{-p} u$, where $p=2/(1-q)$. 
For strict conservation in equation \eqref{scalconslaw2}, 
we need $2p+n-1=0$,
which is seen to imply that $q=(n+3)/(n-1)$ 
is restricted to the conformal power $q_c$. 
(Note, this parallels the condition for scaling invariance of
the action principle for the wave equation.)
From the potential 
\EQ
\Psi_v = G+ x H ,\quad
\Psi_x = \frac{1}{2} (H^2-G^2) -p vH +\frac{k}{q+1} \v{q+1} 
\endEQ
associated with the resulting conservation law \eqref{scalconslaw2}, 
we find by ansatz \eqref{potansatz}
\EQ
\Psi = \frac{k}{2} \frac{q-1}{q+1} x \v{q+1}
\endEQ
and this gives a solution
\EQs
&&
G = p \frac{xv}{x^2-1} \Big( 1- \frac{k}{4} (q-1)^2 \v{q-1} 
\nonumber\\&&\qquad
\pm \sqrt{ (\frac{k}{4} (q-1)^2 x \v{q-1})^2 
- \frac{k}{4} (q-1)^2 (x^2+1)\v{q-1} + 1 }\ \Big) , 
\nonumber\\
&&\\
&&
H = -\frac{k}{2} (q-1) \v{q} + x^{-1} G ,\quad
q=\frac{n+3}{n-1} ,\quad p=\frac{1-n}{2} . 
\label{scalpotsol}
\endEQs

Our results are summarized as follows. 

Proposition~7:
By means of potentials introduced through conservation laws,
the group-resolving systems associated with 
translations, scalings, and inversions 
have the additional solutions 
\eqref{transpotsol}, \eqref{scalpotsol}, \eqref{inverpotsol}, 
for $n>1$. 

\subsection{ Optimal group-resolving systems }

So far we have considered the group foliation resolving systems
that arise from the geometrical subgroups of 
scaling symmetries, translation symmetries, 
and (in the conformally-invariant case $q=q_c$) inversion symmetries
admitted by wave equation \eqref{waveeq}. 
In the case of the conformal power, 
there is an essentially different group foliation 
and associated resolving system given by a subgroup consisting of 
translation symmetries combined with inversion symmetries. 
Note, in general the criterion for foliations to be equivalent
with respect to the full admitted symmetry group is that 
they must come from subgroups related by conjugation. 
For the group-theoretical reasons discussed earlier 
in connection with classical invariant solutions
(cf. \secref{groupaction}), 
an enumeration of conjugacy classes 
under the full group of geometrical point symmetries
of the wave equation \eqref{waveeq} 
is represented by the symmetry generators $\X{scal}$, $\X{trans}$,
in addition to $\X{inver}+\X{trans}$ in the conformally-invariant case. 

To proceed in the latter case, 
the invariants and differential invariants of the point symmetry 
$\X{inver}+\X{trans}$ are straightforward generalizations of 
those derived for inversions. 
We have 
\EQ
x=(1+t^2-r^2)/r ,\quad
v=r^{-p} u ,
\label{invertransxv}
\endEQ
and 
\EQs
G= r^{-p} ( (1+t^2+r^2) \deru{t} +2rt \deru{r} -2p tu ) , 
\label{invertransG}\\
H= r^{-p} ( (1+t^2+r^2) \deru{r} +2rt \deru{t} 
-pr^{-1}(1+ t^2+r^2) u ) ,
\label{invertransH}
\endEQs
where $p=(1-n)/2=2/(1-q)$ and $q=(n+3)/(n-1)=q_c$. 
Through the reciprocal relations
\EQs
\deru{t}= r^{p} (1+(t^2 -r^2)^2+2(t^2+r^2))^{-1}( (1+ t^2+r^2) G -2rt H ) , 
\label{invertransut}\\
\deru{r}= r^{p} (1+(t^2 -r^2)^2+2(t^2+r^2))^{-1}( (1+ t^2+r^2) H -2rt G ) 
+p r^{-1} u , 
\label{invertransur}
\endEQs
we obtain the first-order PDE system 
\EQs
(4+x)^2\xder{G} +G\vder{H} -H\vder{G} =0 , 
\label{invertransGHeq1}\\
(4+x^2)^{-1} ( G\vder{G} -H\vder{H} ) +\xder{H} = -p(p+1) v + k\v{q} , 
\label{invertransGHeq2}
\endEQs
which differs from the conformal-group resolving system 
\sysref{inverGHeq1}{inverGHeq2}
only by the factor $x^2+4$ in place of $x^2$. 

As before, we first try separation of variables ansatzes 
\eqrefs{GHansatz}{GHtwotermansatz}
to solve the group-resolving system \sysref{invertransGHeq1}{invertransGHeq2}.
However, unlike the conformal group case, 
this yields no solutions here. 
Next we try exploiting the conservation law admitted by each of the PDEs 
\eqrefs{invertransGHeq1}{invertransGHeq2},
\EQs
( (x^2+4)^{-1} H/G )_v = ( 1/G )_x , 
\label{invertransconslaw1}\\
( (x^2+4)^{-1} (G^2-H^2) )_v = -2( H +p(p+1) xv +k x \v{q} )_x .
\label{invertransconslaw2}
\endEQs
Through the potential 
$\Phi_x = H/(x^2 G)$, $\Phi_v = 1/G$, 
associated with conservation law \eqref{invertransconslaw1},
no solutions are found 
via the ansatz \eqref{potansatz} or its two-term generalization. 
Using the potential 
$\Psi_x = (G^2-H^2)/(x^2+4)$, 
$\Psi_v = -2H-2x( p(p+1) v +k\v{q} )$, 
from conservation law \eqref{invertransconslaw2}, 
finally we find a two-term ansatz \eqref{pottwotermansatz}
leads to solutions 
analogously to the case of the conformal-group resolving potential system. 

Proposition~8:
Group-resolving system \sysref{invertransGHeq1}{invertransGHeq2}
has the solutions for $n>1$
\EQs
\Psi = kx \v{q+1} -p^2 x \v{2} , 
\\
G =\pm x\sqrt{ (4+k p^{-2} x^2 \v{q-1})(k x\v{q+1} -p^2 x^2 \v{2}) } ,\quad
H = p x v - k p^{-1} x\v{q} ,
\label{invertranspotsol1}
\endEQs
and
\EQs
\Psi = \frac{2k}{q+1} x \v{q+1} -p^2 x \v{2} , 
\\
G =\pm x\sqrt{ 2k (q+1)^{-1} (4+x^2) \v{q+1} -4p^2 \v{2} } ,\quad
H = p x v .
\label{invertranspotsol2}
\endEQs

\section{ Exact solutions } 
\label{results}

To construct solutions $u(t,r)$ of wave equation \eqref{waveeq}
from solutions $G(x,v)$, $H(x,v)$ 
of its symmetry group-resolving systems,
we integrate the corresponding group-invariant equations that define
the symmetry invariants $x,v$, and differential invariants $G,H$
used in formulating those systems in terms of 
$t,r,u,\deru{t},\deru{r}$.
These group-invariant equations
\eqref{scalGH}, \eqref{transGH}, \sysref{inverut}{inverur}, 
\sysref{invertransut}{invertransur} 
take the form of linear, inhomogeneous systems of two first-order PDEs,
in which $G(x,v)$, $H(x,v)$ are explicit expressions.
Because of the inherent symmetry possessed by such systems, 
they reduce to a parametric pair of first-order ODEs
whose integration to obtain $u(t,r)$ is straightforward. 
We will now list all the explicit solution expressions found for $u(t,r)$. 

Theorem~1:
Wave equation \eqref{waveeq} has the following exact solutions 
arising from the explicit solutions of 
its group-resolving systems obtained in Propositions~5 to~8 for $n>1$:
\EQs
&&
u= \left( \pm \sqrt{\frac{k}{2(q+1)}} (q-1) (t+c) \right)^{2/(1-q)} ,
\quad
q\neq -1 , 
\label{usol1}\\
&&
u= \left( \frac{k(q-1)^2}{2(q(1-n)+n+1)} ((t+c)^2-r^2) \right)^{1/(1-q)} ,
\quad
q\neq \frac{n+1}{n-1} ,\quad
n\neq 1 , 
\label{usol2}\\
&&
u= \left( \pm \sqrt{-k} \frac{n-3}{\sqrt{n-2}^3} r 
+ c r^{3-n} \right)^{(n-2)/(n-3)} ,
\quad
q=\frac{4-n}{n-2} ,\quad 
n\neq 2,3 
\label{usol3}\\
&&
u = \left( \frac{k}{(n-2)(n-3)} (c\pm t-r)r \right)^{2-n} ,
\quad
q=\frac{n-1}{n-2} ,\quad
n\neq 2,3 , 
\label{usol4}\\
&&
u =\sqrt{ \pm 2\sqrt{-k} t(1+c t) } ,\quad
u =-\sqrt{ \pm 2\sqrt{-k} t(1+c t) } ,\quad
q=-3 , 
\label{usol5}\\
&&
u= \left( \pm2\sqrt{\frac{k}{n^2-1}} t +c (t^2-r^2) \right)^{(1-n)/2} ,
\quad
q = \frac{n+3}{n-1} ,\quad n\neq 1 , 
\label{usol6}\\
&&
u= \left( \frac{4k}{(n-1)^2} (r^2-t^2)( 1 +2c t +c^2 (t^2-r^2) ) 
\right)^{(1-n)/4} ,
\quad
q = \frac{n+3}{n-1} ,\quad n\neq 1 , 
\label{usol7}\\
&&
u= \left( \frac{4k}{(n-1)^2} 
\Big( t^2 \pm \frac{1}{4} ( c^{-1} \mp c(t^2-r^2) )^2 \Big) 
\right)^{(1-n)/4} ,
\quad
q = \frac{n+3}{n-1} ,\quad n\neq 1 ,
\label{usol8}
\endEQs
where $c$ is an arbitrary constant.

Solutions \eqsref{usol5}{usol8} can be slightly generalized 
by applying a time-translation to $t$. 
It is easy to verify that all the resulting families of solutions, 
with one exception, 
are invariant with respect to 
the full geometrical symmetry group 
admitted by the wave equation \eqref{waveeq},
depending on the nonlinearity power $q$. 
This group comprises translations, scalings, and inversions
for solutions \eqrefs{usol6}{usol7} 
involving the conformal power $q_c=(n+3)/(n-1)$,
and otherwise only translations and scalings
for solutions \eqsref{usol1}{usol5}
involving other powers $q\neq q_c$. 
The solution \eqref{usol8} is exceptional 
as it fails to be invariant under inversions,
despite falling within the conformal power case. 
We note these invariance properties are determined by 
the group-invariant variables of the PDE systems 
from which the individual solutions arise. 
In particular, the lack of conformal invariance
for solution \eqref{usol8} is accounted for
because it comes from the scaling-group resolving system 
rather than the conformal-group resolving system. 
Thus, this solution can be generalized to a two-parameter family
under inversion transformations \eqref{invergroup}, 
yielding
\EQ
u= \left( \frac{4k}{(n-1)^2} 
\Big( \pm r^2 + \frac{1}{4} \Big( c(t^2-r^2) 
\pm c^{-1}(1+2{\tilde c} t +{\tilde c}^2 (t^2-r^2) ) \Big)^2 
\Big) \right)^{(1-n)/4} 
\label{usol9}
\endEQ
with arbitrary constants $c,\tilde c$. 

We remark that, once translations and scalings are taken into account,
the solution families \eqrefs{usol6}{usol7} coming from 
the conformal-group resolving system in Propositions~6 and~7
are the same as the ones obtained through 
the translation-inversion group resolving system in Proposition~8. 

For comparison with classical symmetry reduction, 
note that solution \eqref{usol3} is of translation-invariant form,
and solutions \eqref{usol1}, \eqref{usol2}, \eqref{usol4} are of 
scaling-invariant form to within a time translation, 
as also are solutions \eqrefs{usol6}{usol7} 
to within an inversion transformation. 
Of interest is that the two solutions \eqrefs{usol5}{usol8}
do not have such group-invariant forms 
with respect to any of the geometrical point symmetries 
of wave equation \eqref{waveeq}
(namely, under the action of the full group of geometrical symmetries,
they do not lie on the orbits of any 
scaling-invariant, translation-invariant, or inversion-invariant solutions).

\subsection{ Analytical behavior }

It is certainly noteworthy here that we have found explicit solutions
\eqref{usol3} to \eqref{usol5} 
for three special nonlinearity powers, 
$q=\frac{4-n}{n-2}$, $q=\frac{n-1}{n-2}$, $q=-3$, 
of which the last two are not distinguished by 
any special cases of the symmetry structure of the wave equation 
or of the ODEs for group-invariant solutions, for $n>1$. 
All these powers are subcritical, \ie/ $q <q_*$ 
where $q_*=(n+2)/(n-2)$ is the critical power. 
Two other solutions \eqrefs{usol1}{usol2} hold for almost all powers $q$,
including in particular the critical power 
as well as all supercritical powers.
However, 
the analytical behavior of these two solutions is not so noteworthy,
as $u$ blows up at $t=0$ in solution \eqref{usol1}
and at the light cone $|t|=r$ in solution \eqref{usol2}. 
Related to this blow up, 
the initial data of both solutions at earlier times, $t=t_0<0$, 
has infinite energy whenever $q\ge q_*$, as seen from 
the energy integral
\EQ\label{energy}
\mathcal{E} =\int_0^\infty\Big( 
\frac{1}{2}\deru{t}{}^2 +\frac{1}{2}\deru{r}{}^2 
- \frac{k}{q+1} \u{q+1} \Big)\Bigm|_{t=t_0} r^{n-1} \d r .
\endEQ
Note $\mathcal{E}$ will be conserved, $d\mathcal{E}/dt=0$, 
for any solutions $u(t,r)$ with smooth initial data having suitable 
asymptotic decay for large $r$. 

The analytical features of the solutions with subcritical powers
are worth commenting on. 
First, 
solution \eqref{usol5} is radially homogeneous (\ie/ $r$ independent) 
and hence it satisfies the nonlinear oscillator ODE $\deru{tt}=k u^{-3}$. 
There is no blow up of $u$, but $\deru{t}$ becomes singular 
at $t=0$ as well as $t=-c^{-1}$ in this solution,
which is caused by the negative nonlinearity power. 
Next, 
solution \eqref{usol3} is static (\ie/ $t$ independent).
Asymptotically it is badly behaved for large $r$
since $u$ does not decay. 
Also it is singular at $r=0$ for $c\neq 0$,
so this solution thus describes a monopole with infinite energy. 
Note these two solutions are real-valued in the case $k=-1$,
precisely when the energy integral \eqref{energy} is positive definite. 
The remaining non-static solutions \eqsref{usol4}{usol8} 
with subcritical powers $q < q_*$ 
all exhibit a more interesting blow up of $u$. 

We recall that the blow-up set of a solution $u(t,r)$ 
is defined to be those points in the $(r,t)$ half-plane where 
$u$ or a derivative of $u$ is singular. 
For solutions \eqrefs{usol4}{usol7},
the blow-up sets are given by the light cone $r=|t|$,
in addition to, respectively, 
the timelike line $r=0$ 
and the shifted light cone $r=|t +c^{-1}|$, $c\neq 0$. 
The blow-up set for solution \eqref{usol8} in the ``$+$'' case 
is a pair of shifted light cones 
$r=|t \pm c^{-1}|$, $c\neq 0$. 
Because all these blow-up sets contain a light cone,
the solutions describe initial data with a singularity at
$r_0= |t_0 +c^{-1}| >0$
that propagates forward with unit speed in the $(r,t)$ half-plane. 
This propagation reflects the fact that the wave equation \eqref{waveeq}
is a semilinear hyperbolic PDE with its principal (second derivative) part
given by the radial wave operator $\nder{t}{2}-\nder{r}{2}$
whose characteristic curves are light cones. 

Of the most interest analytically is 
solution \eqref{usol6} with the conformal power $q=q_c$. 
For $k=+1$ and $c<0$ this solution is real-valued. 
Its blow-up set is a hyperbola $|t+t^*|=\sqrt{r^2+t^*{}^2}$,
where $t^*=\frac{\sqrt{k}}{n^2-1} |c|^{-1}$,
which is seen to have a timelike future branch starting at $t=0$
and a spacelike past branch ending at $t=-2 t^*$. 
At intermediate times $0>t>-2 t^*$, there is no blow up for all $r\geq 0$,
and so $u$ is globally smooth in this strip in the $(r,t)$ half-plane.
Consequently, for any fixed time $t=t_0$, with $0>t_0>-2 t^*$, 
we have smooth initial data that evolves to a blow-up at time $t=0$,
as illustrated in Figure~1. 
This initial data, moreover, is found to have finite energy,
due to the asymptotic decay 
$u= |c|^{(1-n)/2} r^{1-n} +O(r^{-n})$ for large $r$. 
But, because of the sign of $k=+1$, 
the energy associated with solutions is not positive definite,
which is what allows the blow-up to occur. 

\begin{figure}
\begin{center}
\includegraphics[scale=0.5]{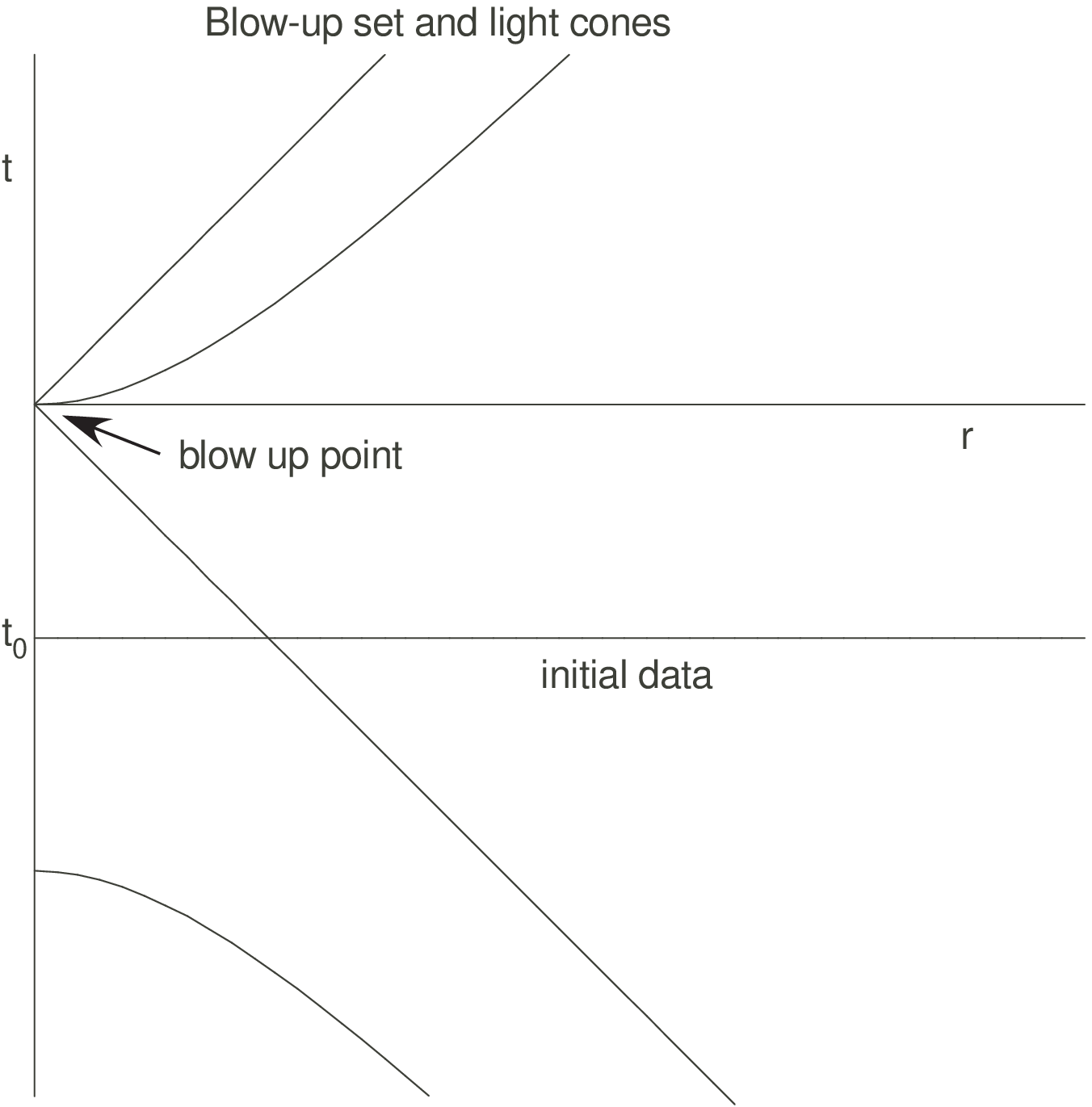}
\end{center}
\caption{}
\end{figure}

Thus, this solution \eqref{usol6} describes
a finite-time blow up for the conformal wave equation
\EQ\label{conformalwaveeq}
\deru{tt} -\deru{rr} -(n-1)\deru{r}/r = k \u{1+2/(n-1)} ,\quad
k=+1
\endEQ
whose nonlinearity is unstable due to its sign. 
The blow-up rate is 
\EQs
&&
u = (2|c|t^*)^{(1-n)/2} t^{(1-n)/2} +O(t^{-(1+n)/2})
\nonumber
\endEQs
for $t\rightarrow 0$ at $r=0$ for solution \eqref{usol6}. 
A worthwhile question is to what extent this behavior 
may be an attractor to which other blow-up solutions evolve
as the blow up time is approached.

Also worthy of analytical interest,
the solution \eqref{usol8} in the ``$-$'' case has no blow up
anywhere in the $(r,t)$ half-plane. 
For large $r$ it behaves like 
\EQs
u=(\sqrt{k}|c|/(n-1))^{(1-n)/2} r^{1-n} +O(r^{-n})
\nonumber
\endEQs
and this asymptotic decay is sufficient to ensure that 
its energy is finite. 
Moreover, the initial data for this solution is smooth for $k=+1$
and evolves with a long-time behavior given by 
\EQs
u = (2\sqrt{k}c/(n-1))^{(1-n)/2} t^{1-n} +O(t^{-n})
\nonumber
\endEQs
for large $t$. 
It is thus a global smooth solution of finite energy 
for the conformal wave equation \eqref{conformalwaveeq}. 

The contrasting behaviors of the previous two solutions 
with $k=+1$ for the conformal power $q=q_c$ 
is typical of wave equations with an unstable nonlinearity \cite{Strauss}. 

We remark that solution \eqref{usol9} has a similar analytical behavior
to solution \eqref{usol8}. 

\subsection{ Group-invariant solutions }

Finally, we will also consider for comparison the solutions 
\eqref{transvscalsol}, \eqref{transvdilsol}, \eqref{scalvdilsol}
derived by classical symmetry reduction 
of the wave equation \eqref{waveeq}. 
These solutions were found to satisfy 
nonlinear, unstable oscillator equations, 
involving the special powers
$q=\frac{n+2}{n-2}$, $q=\frac{4-n}{n-2}$
for the nonlinearity. 
It is possible to carry out the quadratures to obtain the solutions
in explicit form in the special case when the integration constant
corresponding to the oscillator energy is zero. 
In terms of the canonical coordinates $x,v$ 
adapted to the symmetry reductions,
the quadratures \eqrefs{scalvdilsol}{transvscalsol} both yield 
\EQ
v= \left( \frac{n(n-2)}{4k}( 1-(\tanh x)^{\pm 2} \right)^{(n-2)/4} , 
\endEQ
while the other quadrature \eqref{transvdilsol} gives
\EQ
v=\left\{ \begin{array}{ll}
\exp( \pm\sqrt{-k} x ) 
&,\mbox{ for $n=3$,}\\
\left( \pm \sqrt{\frac{k}{n-2}} (n-3) x \right)^{(n-2)/(n-3)} 
&,\mbox{ for $n\neq 2,3$.}\end{array} \right.
\endEQ
Hence we obtain scaling-invariant solutions
\EQs
&& 
u= \left( \frac{n(n-2)}{4k} \right)^{(n-2)/4} t^{1-n/2} ,\quad
q=\frac{n+2}{n-2} ,\quad
n\neq 2 , 
\label{invusol1}\\
&& 
u= \left( \frac{n(n-2)}{4k} \right)^{(n-2)/4} (r^2 -t^2)^{(2-n)/4} ,\quad
q=\frac{n+2}{n-2} ,\quad
n\neq 2 , 
\label{invusol2}
\endEQs
and translation-invariant solutions
\EQs
&& 
u= \left( \pm \frac{n(n-2)}{k} \right)^{(n-2)/4} (r^2 \pm 1)^{1-n/2} ,\quad
q=\frac{n+2}{n-2} ,\quad
n\neq 2 , 
\label{invusol3}\\
&& 
u= \left( \pm\frac{n-3}{n-2}\sqrt{\frac{k}{2-n}} \right)^{(n-2)/(n-3)} 
r^{(n-2)/(n-3)} ,\quad
q=\frac{4-n}{n-2} ,\quad
n\neq 2,3 
\label{invusol4}
\endEQs
(we omit the case $n=3$ for the latter solution 
as the power $q=1$ is trivial). 
In addition, we also list the special invariant solutions 
\eqrefs{transvinvscalsol}{invervinvdilsol}
arising from the extra point symmetries admitted by 
the translation, scaling, and inversion symmetry reductions. 
These yield
\EQs
&& 
u= \left( \frac{n-2}{2\sqrt{k}} \right)^{(n-2)/2}
r^{1-n/2} ,\quad
q=\frac{n+2}{n-2} ,\quad
n\neq 2 , 
\label{invusol5}\\
&& 
u= \left( \frac{(n-1)(n-3)}{4k} \right)^{(n-1)/4}
r^{(1-n)/2} ,\quad
q=\frac{n+3}{n-1} ,\quad
n\neq 1 . 
\label{invusol6}
\endEQs

We note solutions \eqref{invusol1}, \eqref{invusol2}, \eqref{invusol4}
are specializations of solutions \eqsref{usol1}{usol3}.
Thus, the only extra solutions arising from symmetry reduction 
are the ones \eqref{invusol3}, \eqrefs{invusol5}{invusol6},
all of which are static. 

The two static solutions \eqrefs{invusol5}{invusol6}
clearly are singular at $r=0$
and hence describe monopoles of wave equation \eqref{waveeq}
with the critical power $q=q_*$ and conformal power $q=q_c$, respectively.
However, it is readily seen that their energy is infinite
due to the asymptotic rate of decay of $u$ with $r$. 

In contrast, the other static solution \eqref{invusol3} 
with critical power $q=q_*$ 
has finite energy and is non-singular at $r=0$.
In the case $k=-1$, 
where the nonlinearity of the wave equation is stable,
this solution is singular at radius $r=1$
(note this location can be moved to any non-zero radius $r=r^*$ 
by applying a scaling symmetry to the solution). 
In the opposite case $k=+1$, 
there is no singularity at any radius $r\geq 0$
and the solution is globally smooth, 
but here the nonlinearity is unstable. 
Thus, this solution \eqref{invusol3} describes
a finite-energy static ``soliton'' 
for the unstable critical-power wave equation
\EQ\label{criticalwaveeq}
\deru{tt} -\deru{rr} -(n-1)\deru{r}/r = k \u{1+4/(n-2)} ,\quad
k=+1 .
\endEQ
The existence of such a solution suggests that there may be 
very interesting analytical long-time behavior 
for smooth finite-energy solutions of this wave equation.

\section{ Remarks }
\label{remarks}

Group foliations in general
convert a nonlinear PDE into an equivalent system of
first-order equations (the group-resolving system) 
by means of group-invariant variables 
associated with an admitted group of transformations.
The construction of group foliations using admitted point symmetry groups
of nonlinear PDEs is originally due to Lie and Vessiot 
and was revived in modern form by Ovsiannikov \cite{Ovsiannikov}. 
Because the equations for a group foliation contain 
all solutions of the given nonlinear PDE, 
ansatzes or differential constraints must be used to reduce the equations
into an overdetermined group-resolving system
to obtain explicit solutions. 
Compared with classical symmetry reduction, 
a main difficulty to-date has been how to find 
effective, systematic ansatzes that lead to useful reductions. 

Recent work \cite{SheftelWinternitz,Sheftel} has introduced 
an ansatz based on imposing commutativity of operators for
group-invariant differentiation,
developed for the example of 
the infinite-dimensional conformal symmetry group
of the ``heavenly equation'' that describes 
the geometry of self-dual gravitational fields in General Relativity. 
This ansatz is shown to give a nontrivial reduction of 
the conformal-group foliation equations, 
yielding explicit exact solutions. 
Other work, on hydrodynamic equations 
with infinite-dimensional symmetry groups,
appears in \Ref{Ovsiannikov,Golovin}.

The semilinear $n$-dimensional radial wave equation 
we have considered is an example with a finite-dimensional symmetry group. 
Our results show that a basic separation of variables ansatz is effective
in reducing the group foliation equations,
especially when combined with the use of potentials based on
conservation laws of these equations. 
It is significant that we find some reductions give 
exact solutions having interesting analytical properties. 

For future work, applications of these methods 
to wave equations with nonlinear wave speeds 
and to nonlinear heat conduction equations,
as well as to integrable nonlinear evolutionary equations 
would be fruitful to consider. 
It is expected that in general our method 
should be well suited to equations with power nonlinearities. 
Concerning group-theoretical aspects, 
it would also be worthwhile to examine ansatzes for 
reduction of group-resolving systems 
in connection with the well-known non-classical method 
\cite{nonclassical1,nonclassical2,nonclassical3}
for finding exact solutions of nonlinear PDEs 
through invariant surface conditions.

\section{Acknowledgement}
S.L. thanks the Mathematics Department of Brock University for support
during the period of a research visit when this paper was written. 
S.A. is grateful to Andrew Copfer for assistance with some calculations
at an early stage of the research. 
The referee is thanked for useful comments which have improved parts of
this paper. 

\appendix
\section{Appendix}

Here we state the form of 
the translation subgroup, scaling subgroup, and inversion subgroup 
comprising the one-dimensional geometric point symmetry groups 
admitted by wave equation \eqref{waveeq}
for nonlinearity powers $q\neq 1$. 
Their action on solutions is described by the following
one-parameter ($\lambda$) transformations:
\EQs
&&\eqtext{translations}\quad
u(t,r) \rightarrow u(t+\lambda,r) 
\quad\eqtext{for all $q$}, 
\label{transgroup}\\
&&\eqtext{scalings}\quad
u(t,r) \rightarrow \lambda^{-p} u(\lambda t,\lambda r) 
\quad\eqtext{for all $q$}, 
\label{scalgroup}\\
&&\eqtext{inversions}\quad
u(t,r) \rightarrow \Omega_{(-\lambda)}^{p} 
u(\Omega_{(\lambda)}(t+\lambda (t^2-r^2),\Omega_{(\lambda)} r)
\quad\eqtext{only for $q=q_c$}, 
\label{invergroup}\\
&&\severalquad\eqtext{with}\quad
\Omega_{(\lambda)}=\frac{t^2-r^2}{(t+\lambda (t^2-r^2))^2 -r^2} , 
\nonumber
\endEQs
where $p=2/(1-q)$. 
We note that, in terms of the invariants
$\xi=t/r$, $x=(t^2 -r^2)/r$, $v=r^{-p} u$
associated with scalings and inversions,
the one-parameter inversion transformations \eqref{invergroup} 
act in a simple way by 
\EQ
v(\xi,x) \rightarrow v(\xi+\lambda x,x)
\endEQ
where $v(\xi,x) = r^{-p} u(t,r)$ for solutions $u(t,r)$. 
Finally, if we combine 
an inversion \eqref{invergroup} for $\lambda=1/\epsilon$
and a scaling \eqref{scalgroup} for $\lambda=\epsilon^2$
together with a translation \eqref{transgroup} for $\lambda=\epsilon$,
the limit $\epsilon \rightarrow 0$ yields
the special conformal transformation \eqref{involution}
belonging to the full point symmetry group $SL(2,\mathbb R)$
in the conformal power case $q=q_c$.
This conformal symmetry is distinguished by providing 
an involution under which the translation and inversion subgroups 
are exchanged while the scaling subgroup is kept invariant.


\begin{thebibliography}{99}
\bibitem{integrable}
As well-known special cases,
linear PDEs under similarity reduction give linear ODEs solved by
special functions, \eg/ 
similarity solutions of the heat equation are 
described by parabolic cylinder functions; 
completely integrable nonlinear evolutionary PDEs, 
such as the KdV equation or sine-Gordon equation, 
invariably yield Painlev\'e ODEs for similarity solutions, 
\eg/ the second Painlev\'e transcendent describes similarity reductions 
of the KdV equation.
\bibitem{orbits}
More precisely, 
each point in the solution jet space of the group-resolving system 
is mapped into a one-dimensional subspace of the original solution jet space.
The subspace either is an invariant family of solutions 
comprising a one-dimensional orbit of the group 
(which is the general case, 
where the mapping of points to orbits is one-to-one),
or is a set of invariant solutions 
each of which is a degenerate orbit of the group
(in which case the mapping is one-to-many).
\end{thebibliography}
\end{document}